\documentclass[final,5p,times,twocolumn]{elsarticle}

\usepackage{amssymb}
\usepackage{amsmath}
\usepackage{hyperref}
\usepackage{booktabs}
\usepackage[utf8]{inputenc}
\usepackage[T1]{fontenc}
\usepackage{lineno}
\usepackage{comment}
\usepackage{caption}
\usepackage{subcaption}
\journal{Nuclear Inst. and Methods in Physics Research, A}

\begin{document}

\begin{frontmatter}



\title{Experience gained about Resistive Plate Chambers ageing from the ALICE Muon TRigger/IDentifier detector}

\author[1,2]{\small A. Ferretti}
\ead{ferretti@to.infn.it}

\author{for the ALICE Collaboration}

\affiliation[2]{organization={Università degli studi di Torino, Dipartimento di Fisica and INFN, Sezione di Torino},
             addressline={Via P. Giuria 1},
             city={Torino},
             postcode={10125},
             state={Italy},
             country={}}

\begin{abstract}
\small The ALICE Muon IDentifier is composed of 72 single-gap bakelite Resistive Plate Chambers, which
have been operational since 2009 in maxi-avalanche mode (discrimination threshold:7 mV without amplification) with a Tetrafluoroethane/Isobutane/Sulfur Hexafluoride gas mixture, undergoing counting rates of the order of tens of Hz/cm$^2$.

\small In this talk, the long-term performance and stability of the RPC system will be discussed, in terms
of efficiency, dark current and dark rate. An assessment of potential signs of ageing observed on
the detectors will be presented, together with a summary of the most common hardware problems
experienced.

\end{abstract}


\end{frontmatter}


\section{Introduction}
\label{sec:intro}

The ALICE experiment at CERN \cite{alice1} studies p-p, p-A and A-A collisions at LHC energies in order to characterize the Quark-Gluon Plasma, a state of nuclear matter reached under extreme conditions of temperature and pressure. The ALICE detector is equipped with a Forward Muon Spectrometer (FMS)\cite{ALICE:1999aa}, covering the pseudorapidity range 2.5<$\eta$<4 and mainly dedicated to measuring charmonia and bottomonia states in their muonic decay channel. The ALICE Muon TRigger system (MTR) is designed to identify muons tracked by the FMS and to perform a transverse momentum cut in order to select high-$p_{\text t}$ muons that traverse the ALICE muon spectrometer.

\begin{figure}[h!]
\center \includegraphics[height=0.6\linewidth]{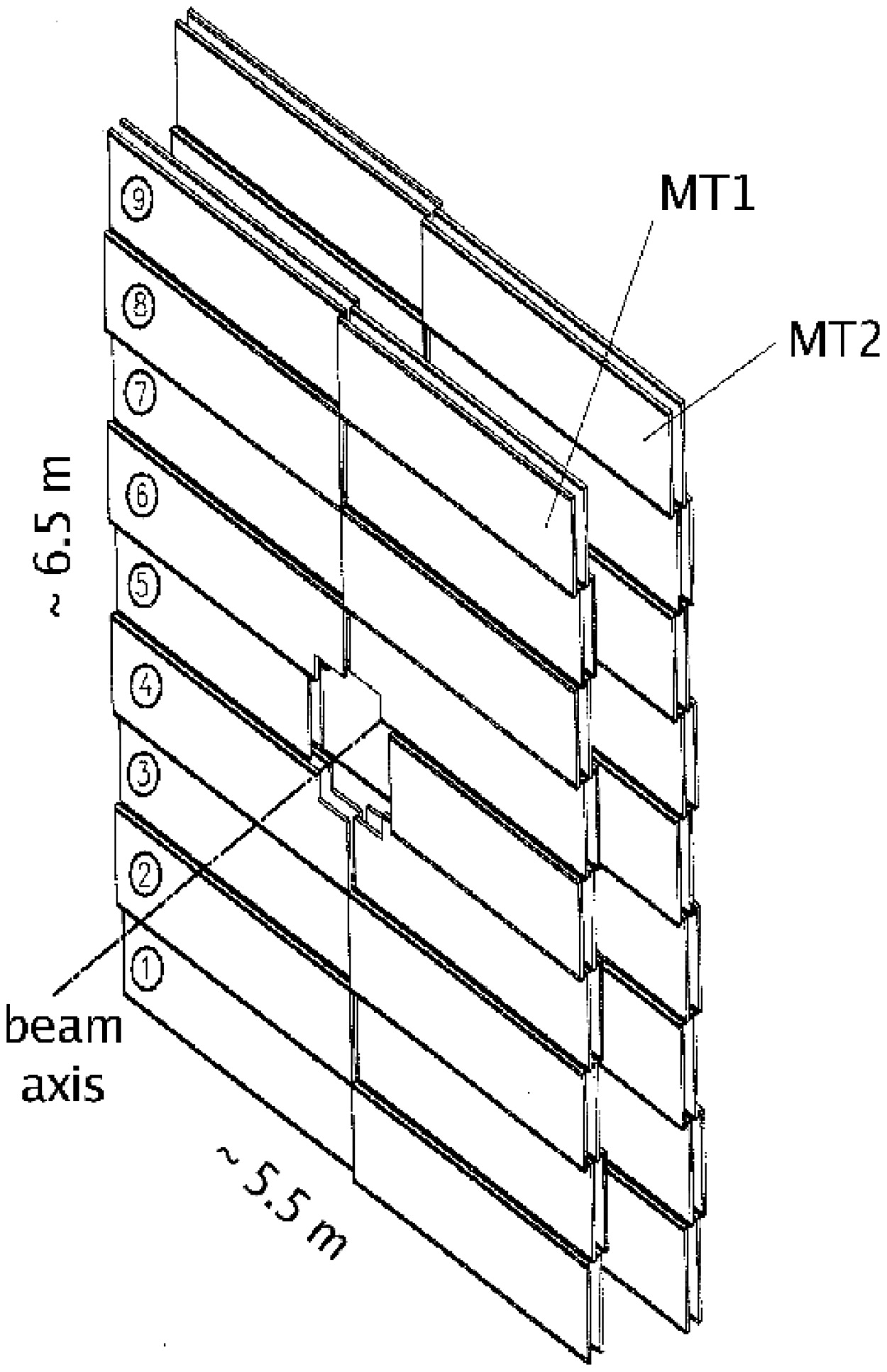}
\caption{Layout of the ALICE Muon TRigger/IDentifier. Each slat is one RPC}
\label{fig:MIDlayout}
\end{figure}

The Muon Trigger\cite{ALICE:1999aa} (renamed since 2020 as Muon IDentifier, MID) is placed downstream of the FMS, starting $\sim$16~m from the Interaction Point and is made of 72 Resistive Plate Chambers (RPCs), arranged in two stations of two planes each (see Figure~\ref{fig:MIDlayout}). Each detection plane is $\sim$5.5~m wide and $\sim$6.5~m tall and it is made of 18 chambers, which come in two sizes: sixteen in each plane are $\sim$270$\times$70 cm$^2$, while the remaining two are shorter ($\sim$210$\times$70 cm$^2$) in order to accommodate the beam pipe shielding in the middle of the plane. MID RPCs are equipped with orthogonal readout strips to provide spatial information along both X and Y directions, for a total of 21k readout channels.

\section{Resistive Plate Chambers of the ALICE Muon Trigger}
\label{sec:RPC}

The RPCs of the ALICE Muon Trigger\cite{RPC1} are single-gap detectors equipped with Bakelite electrodes, which define a gas gap 2~mm thick. Electrodes are 2~mm thick, too, and their resistivity is in the range of $3\cdot$10$^9$-1$\cdot$10$^{10}$ $\Omega$ $\cdot$cm. In the first tests, RPCs were internally coated with a thick linseed oil layer, obtained by filling the gas gap with a mixture of linseed oil and pentane (to dilute the oil). The purpose of the oil layer is to reduce the noise counting rate of the chamber, presumably by smoothing out the micro-irregularities of the internal cathode surface, which could trigger  field electron emission. In our first prototypes, the proportion between oil and pentane was 70/30, resulting in a thick oil layer ($\sim$30~$\mu$m).

Initially, these detectors were designed to operate in streamer mode because simulations foresaw that the large number of particles, which could be created in Pb-Pb collisions in the MTR planes' acceptance could have an impact on trigger selectivity and to prevent this effect, a small cluster size was needed. To this end, a gas mixture made of 49\% Argon, 40\% C$_{2}$H$_{2}$F$_{4}$, 7\% i-C$_{4}$H$_{10}$ and 4\% SF$_{6}$ was devised, and the signal discrimination threshold was set  at 80~mV without any preamplification stage\cite{adult}.

Successful beam tests on 50$\times$50 cm$^2$ prototypes were performed in 1998 and 1999 at CERN\cite{RPC1, spat}. Moreover, to ascertain the rate capability of the detector while uniformly irradiated over all its surface, a one-week test was performed at the CERN Gamma Irradiation Facility by exposing one  prototype to the 667~keV $^{137}$Cs source to induce a single counting rate of 300~Hz/cm$^2$.

\section{First ageing tests}
\label{sec:age}

\begin{figure}[h!]
\center \includegraphics[height=0.55\linewidth]{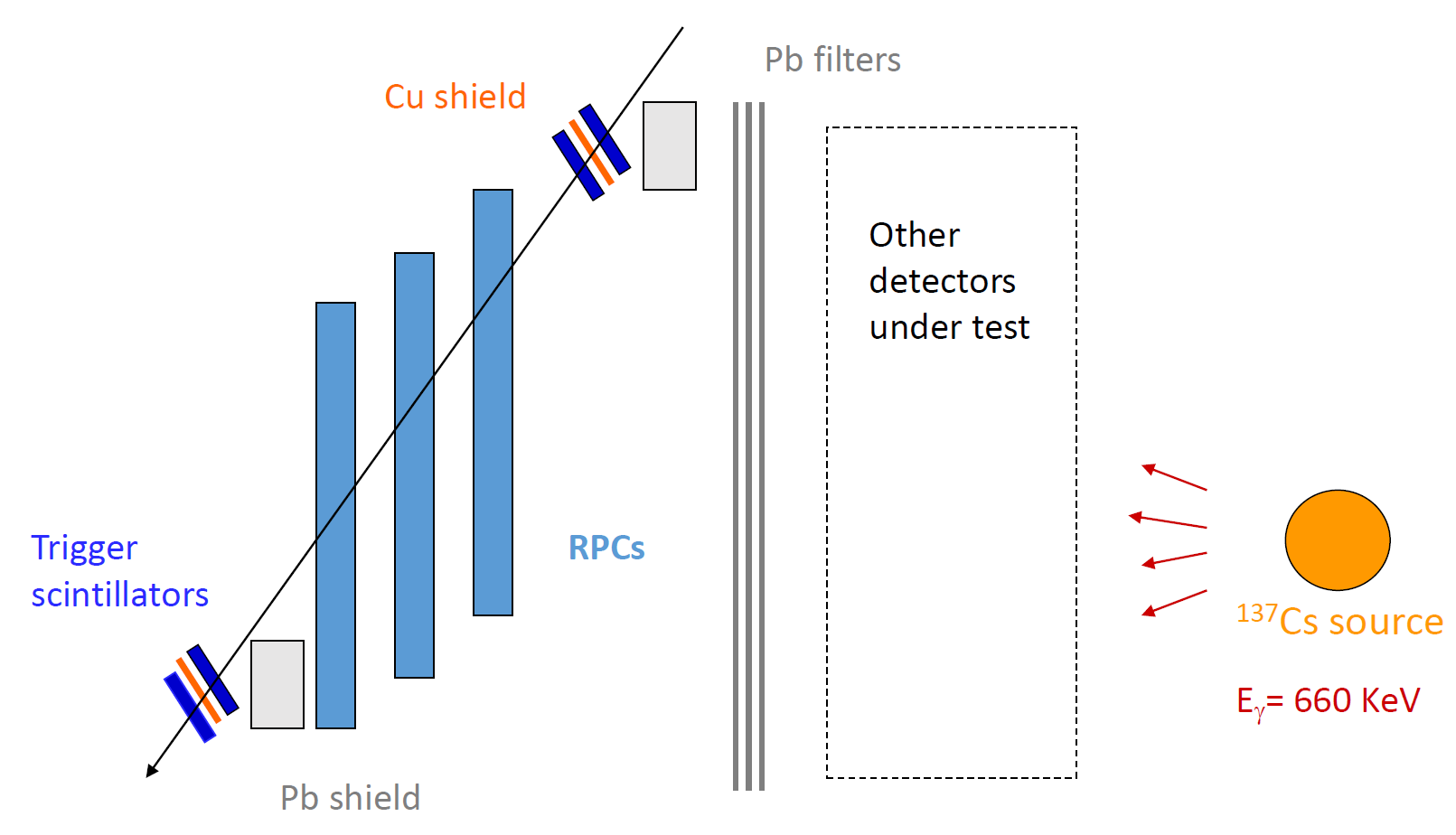}
\caption{Schematic layout of the GIF ageing test}
\label{fig:GIFlayout}
\end{figure}

After successful tests, the Muon Trigger group started an R\&D campaign to evaluate the ageing characteristics of the detectors in order to estimate their lifetime. To this end, in February 2001, we installed two 50$\times$50 cm$^2$ prototypes in the GIF, and we exposed it to a $\gamma$-induced counting rate in the 70-150~Hz/cm$^2$ range\cite{age1}. Efficiency was monitored on a small area ($\sim$30$\times$10 cm$^2$) using a scintillator telescope selecting cosmic rays crossing the upper-central zone of the detectors (see Figure~\ref{fig:GIFlayout}). Unlike the RPCs used in previous tests, one of these prototypes was coated with a much thinner layer of linseed oil because of the experience of the BaBar IFR RPCs\cite{babar}, which underwent major efficiency problems due to an excess of linseed oil. The proportions between oil and pentane were then changed to 30/70, resulting in a oil layer $\sim$3~$\mu$m thick. The other RPC was without oil coating at all.

After one month of irradiation, corresponding to an integrated charge of 30~mC/cm$^2$, a large increase of both working current and dark current was observed in the oiled RPC. The working current increased by $\sim$40~$\mu$A, and the same increase was measured for the dark current (i.e. the current drawn by the chamber operated at the HV working point when the $\gamma$ source is off), which went from 2~$\mu$A to more than 40~$\mu$A. The dark hit rate, (i.e. the hit rate measured operated at the HV working point when the $\gamma$ source is off) went from less than 0.1~Hz/cm$^2$ to $\sim$3.5 Hz/cm$^2$. The non-oiled chamber fared even worse: after integrating a charge as low as 5~mC/cm$^2$ the dark current went from 2.5~$\mu$A to more than 200~$\mu$A and the dark rate increased from 1.2 Hz/cm$^2$ to 8.4~Hz/cm$^2$.

These results were unexpected because, in the above-mentioned GIF test in 1999, we did not observe any similar increase in dark current and/or rate. We stopped the test and opened two RPCs in order to check the inside of the gas gap, and we discovered that the internal surface had deteriorated. Chemical analysis performed at CERN\cite{age2} showed large quantities of Fluorine over all the internal area, while in a unused detector no Fluorine at all was found. 

In order to better understand what was happening, we started another ageing test at GIF in May, 2001 with the same setup as above\cite{age2}. 
 We suspected that the large amount of SF$_6$ in the mixture could be linked to the current increase; thus, we assumed that the oil layer thickness was playing a critical role, so we decided to perform two separate comparison tests. In the first one, we compared the behaviour of two identical prototypes, one flowed with the original streamer mixture and the other flowed with an identical mixture but with the SF$_6$ percentage set to 1\% (i.e. 50.5\% Argon, 41.3\% C$_{2}$H$_{2}$F$_{4}$, 7.2\% i-C$_{4}$H$_{10}$, 1\% SF$_{6}$). Since SF$_6$ is a strong quencher, the HV of the RPC with 1\% of SF$_6$ was lowered by 1~kV w.r.t. the one with 4\% in order to operate the detector at the voltage corresponding to the new efficiency knee. In the second test, we compared two identical prototypes, but for the thickness of the oil layer: on one of the two, the oiling process was performed twice in order to get a thicker layer. To compare the ageing we measured the dark current as a function of the total current integrated by the chambers per area unit. 

\begin{figure}[h!]
\center \includegraphics[height=0.6\linewidth]{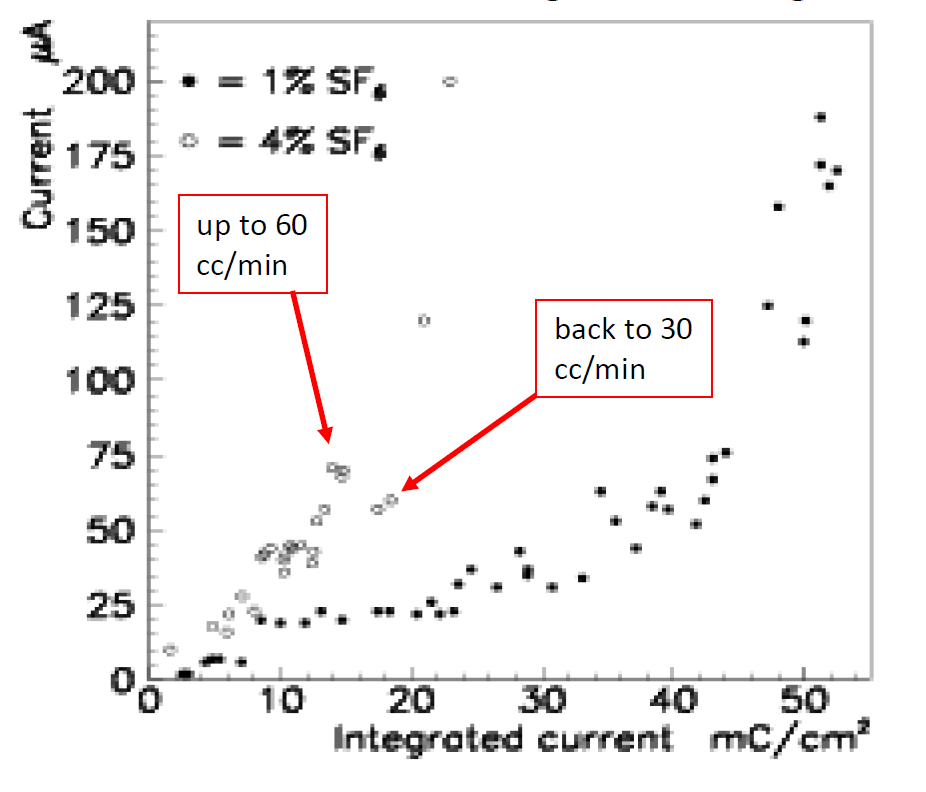}
\caption{Dark current drawn by RPCs as a function of integrated current. Black dots refer to an RPC flowing with a mixture of  1\% SF$_6$, while white dots refer to a RPC flowing with a mixture of 4\% SF$_6$ The temporary decrease of the dark current in the RPC with 1\% SF$_6$ is correlated with an increase in the gas flow rate.}
\label{fig:4vs1}
\end{figure}

Results of the first test (see Figure~\ref{fig:4vs1}) showed that operating the gas mixture with 1\% of SF$_6$ caused a strong reduction of the dark current increase: even if the chamber flowed with 1\% was exposed to a counting rate double of the one of the 4\% chamber, (100~Hz/cm$^2$ vs. 50~Hz/cm$^2$), the dark current stayed under 200~$\mu$A up to 60~mC/cm$^2$, while the 4\% chamber reached the same value before integrating 30~mC/cm$^2$. 

\begin{figure}[h!]
\center \includegraphics[height=0.6\linewidth]{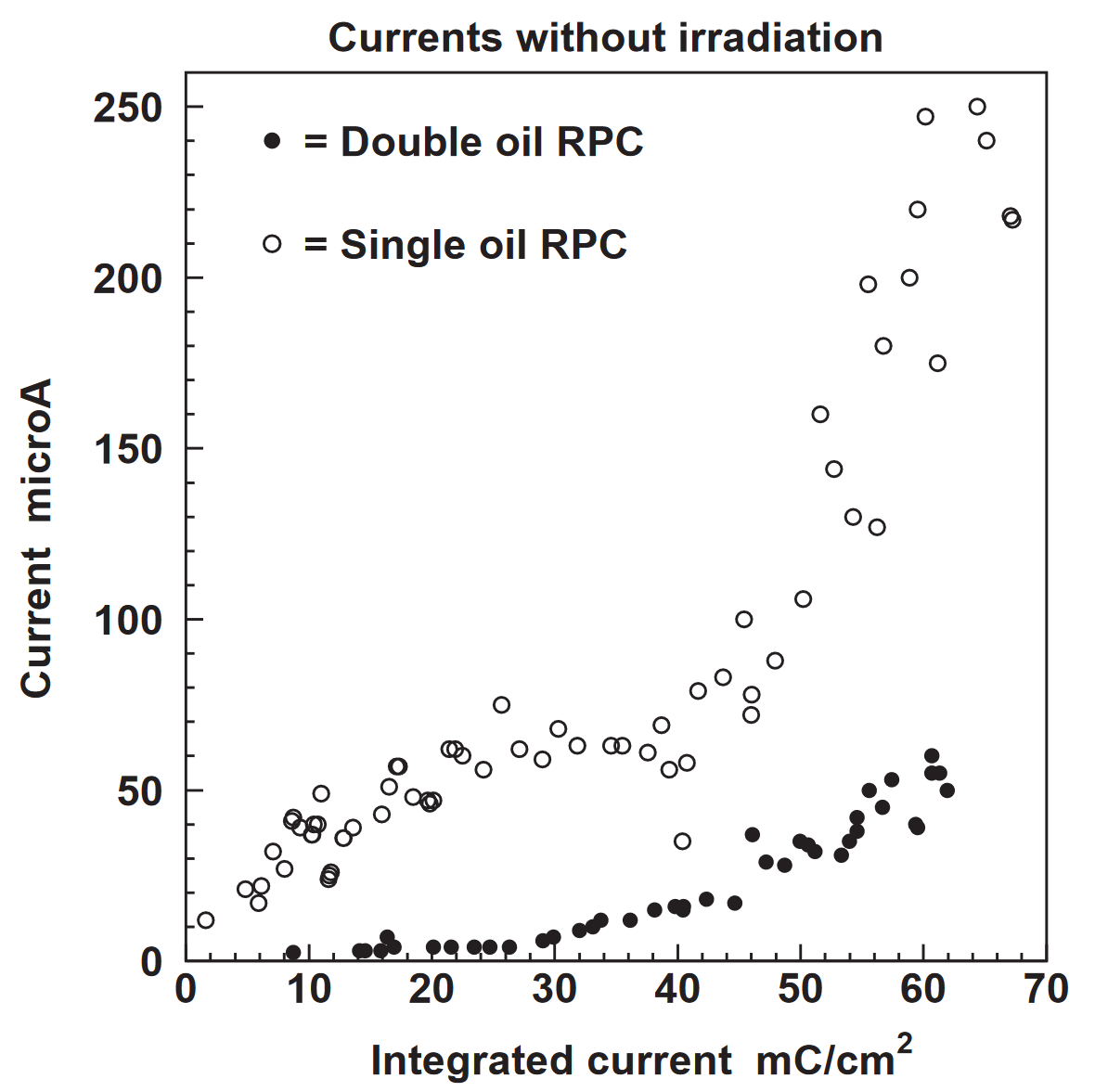}
\caption{Dark current drawn by RPCs as a function of integrated current. Black dots refer to an RPC internally coated with a double linseed oil layer, while   white dots refer to an RPC coated with only one oil layer.}
\label{fig:2Lvs1L}
\end{figure}

After this outcome, for the second comparison, we flowed the RPCs with the ``1\% SF$_6$'' mixture and the results showed that the RPC with a double oil layer performed better than the one with a single layer. After integrating a current of 60~mC/cm$^2$, the ``double oiled'' RPC absorbed a dark current of about 50~$\mu$A (see Figure~\ref{fig:2Lvs1L}) while the ``single oiled'' one reached values above 200~$\mu$A. A test with chambers oiled thrice did not show significant improvements with respect to the twice-oiled' one\cite{age2}.

\section{Long ageing tests}
\label{sec:longage}

\begin{figure}[h!]
\center \includegraphics[height=0.6\linewidth]{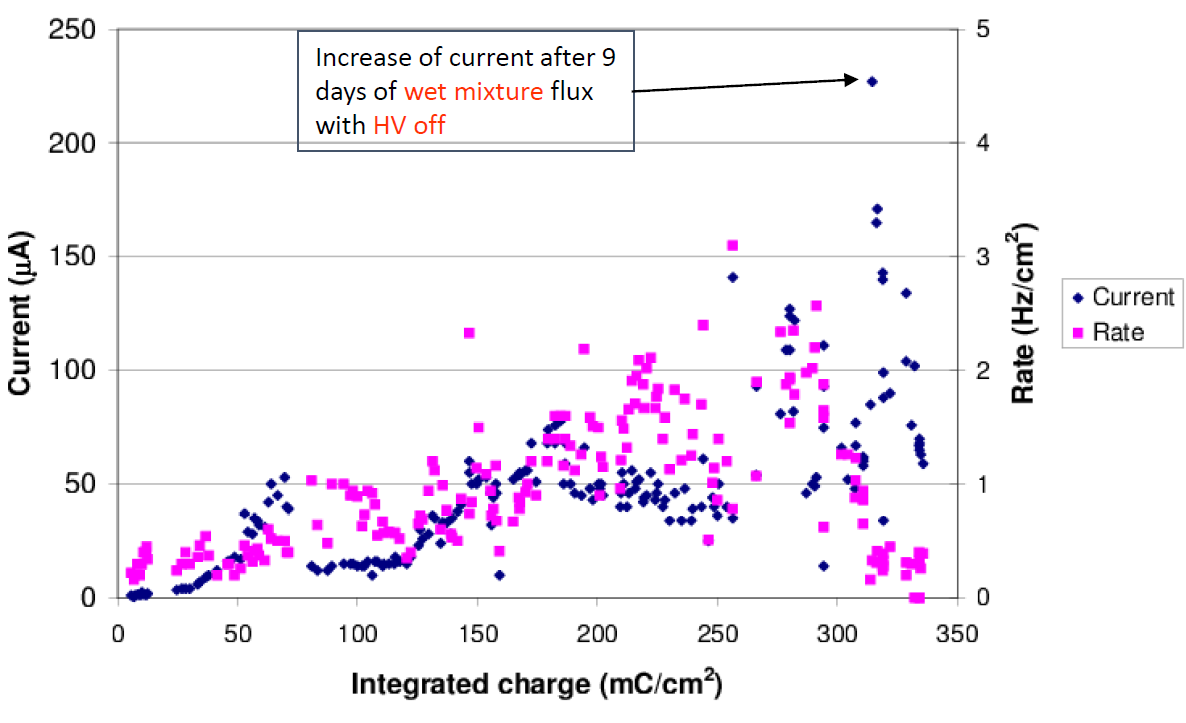}
\caption{Dark current (lozenges, left vertical axis) and dark counting rate (squares, right vertical axis) drawn by an RPC as a function of integrated current.}
\label{fig:Longgif}
\end{figure}

The outcome of the tests reported above led us to a year-long test at GIF on an RPC with a double oil layer and flowing with 1\% SF$_6$ streamer mixture, with the same setup mentioned above (with the addition of scalers to measure the dark hit rate), in order to ascertain if the RPC was able to withstand an integrated current comparable with the one required in the first 10 years of ALICE data-taking. The efficiency, measured only partially on the detector's active area, stayed at 90\% until the end of the test (i.e. after 340~mC/cm$^2$). As shown in Figure~\ref{fig:Longgif}, the dark current remained around 50~$\mu$A up to an integrated charge of 250~mC/cm$^2$ (with drops observed after GIF source shut-off periods) while the dark rate remained under 100~Hz/cm$^2$. Then, the dark current started to increase while the dark rate started to decrease. 

\begin{figure}[h!]
\center \includegraphics[height=0.6\linewidth]{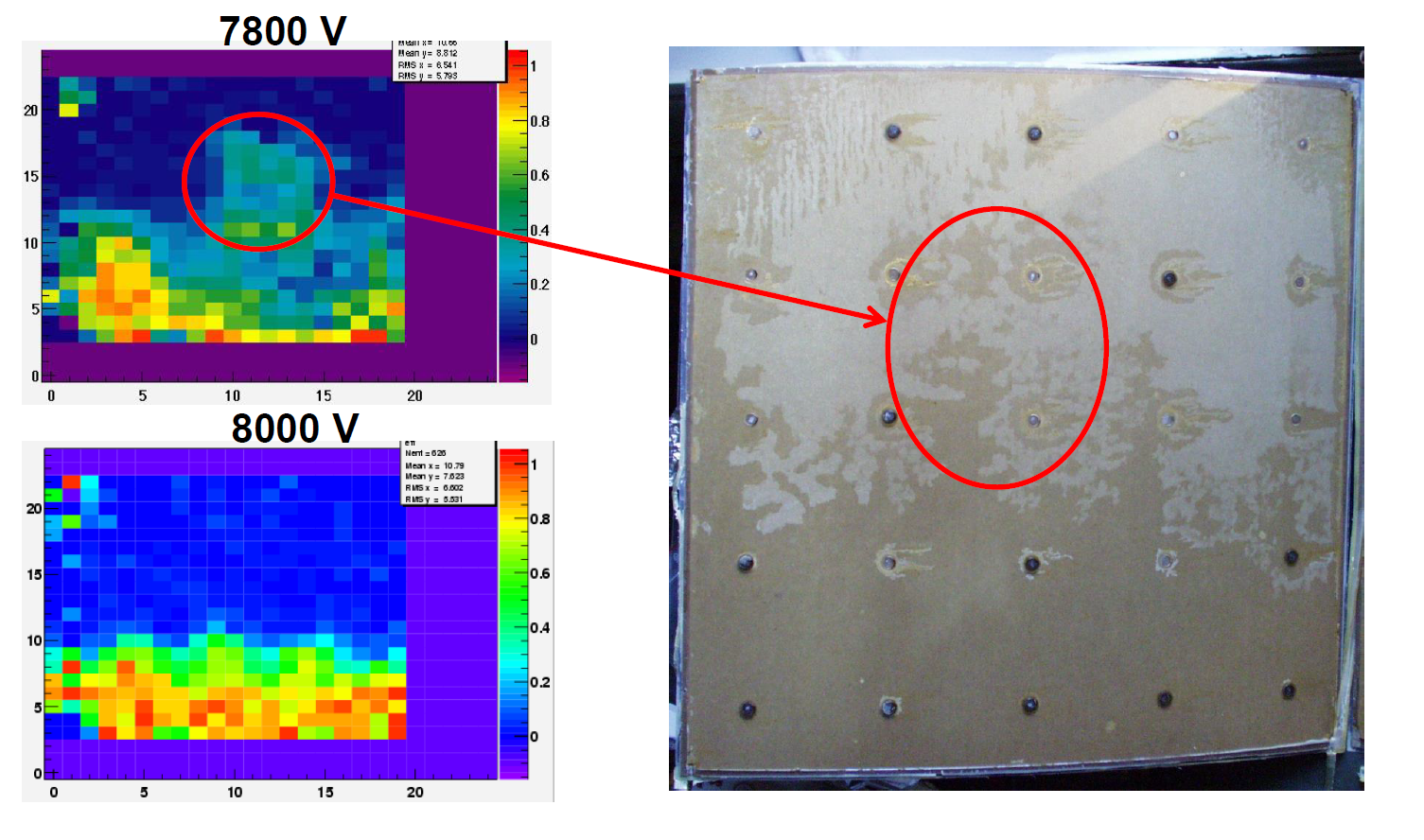}
\caption{Top left: map of efficiency of an aged 50$\times$50~cm$^2$ RPC working with an H.V. of 7.8~kV. Each small square represents a $2.5\times2.5$cm$^2$ detector area. Bottom left: same as top, but with an H.V. of 8.0~kV. Right: picture of the internal surface of the aged chamber. Brighter areas are the ones where the linseed oil layer has been removed.}
\label{fig:eff}
\end{figure}

This behaviour was puzzling because, before that point, we thought the dark current was correlated to the dark hits. We decided to stop the test and study the efficiency of the RPC all over the surface with the help of the Cosmic Rays Tracking Station in the INFN Torino lab and discovered (Figure~\ref{fig:eff}, left) that a large part of the detector outside the area in which we measured the efficiency at GIF was inefficient, and that in the inefficient zone the RPC single rate was zero: the behaviour of the detector was then similar to the one when it is filled with Argon and operated above 2.5~kV, i.e. large currents but with zero hit rate.

Thanks to this test, a hypothesis arose: the dark current is due to electrons emitted from localized areas of the internal surface of the cathode (corresponding to the inefficiency zones) at a high rate, i.e. higher than the rate capability of the detector).
This could have explained some observed facts:
1) The dark current is zero when the HV of the chamber is under the threshold where electron multiplication starts, and it starts increasing
2) Increasing the voltage ends up in a dark current increase, that is limited only by the resistivity of the electrodes in a way similar when the RPC is filled with Argon.  Then, over a certain voltage (which depends on rate of primary noise electrons, electron multiplication and electrode resistivity) the current becomes continuous and does not induce pulsed signals on the readout strips.
3) The chamber is inefficient in the areas interested in the phenomenon of cathode electron emission.
4) According to Ohm's law, the large current through the electrodes causes a drop in the effective voltage applied to the gas gap, which, in turn, explains the observed decrease in efficiency when increasing voltage.

To understand the cause of the hypothesized electron emission, we opened the detector in order to examine the internal electrode surface and we discovered that in the inefficient zones, the linseed oil layer was no longer there (Figure~\ref{fig:eff}, right). Since the oil layer's role is to smooth out the cathode surface to minimize  field electron emission, it is plausible that the removal of the layer would trigger a large emission of electrons from the cathode with the consequences mentioned above. But why the linseed oil disappeared? A possible explanation was that the removal was due to the production of chemically active fragments from collisions between the fluorinated gas molecules present in the gas mixture and the avalanche electrons, which then reacted with the oil layer corroding it.

\section{Exhaust gas analysis}
\label{sec:exhaust}

In order to understand the causes of the oil coating loss, in 2003 we started an R\&D campaign in the framework of the ``Common Effort for RPCs'' project, in collaboration with the CERN EST division and of the CERN Gas Group. The testing procedure was as follows: RPC prototypes were operated at GIF under irradiation with different gas mixtures, and the exhaust gas was bottled and chemically analyzed utilizing gas chromatography in order to measure the amount and identify the impurities in the exhaust gas mixture generated by RPC operation.

The first chemical analyses showed that corrosive hydrofluoric acid (HF) was present in the exhaust mixture and that its concentration was roughly proportional to the operating current of the RPCs. In order to understand how much HF was due to the presence of C$_{2}$H$_{2}$F$_{4}$ and how much was coming from SF$_{6}$, in April 2004 we operated one 50$\times$50 cm$^2$ RPC with the standard gas mixture 50.5\% Ar, 41.3\% C$_{2}$H$_{2}$F$_{4}$, 7.2\%i-C$_{4}$H$_{10}$ and 1\%SF$_{6}$ (mix \#1 in table\ref{tab:table1}) and we compared its exhaust gas analysis with the results obtained from an identical RPC operated with a gas mixture made of 80\% Ar and 20\% i-C$_{4}$H$_{10}$, with the addition of 1\% SF$_{6}$  (mix  \#2 in Table\ref{tab:table1})
The detector flowed with the mixture without C$_{2}$H$_{2}$F$_{4}$ showed signs of ageing quite soon, and when the gas analysis was performed it was drawing a dark current of 90~$\mu$A.

The comparison between the chemical analyses of the two mixtures can be seen in the two first rows in Table~\ref{tab:table1}.

\begin{table}[h!]
  \begin{center}
    \caption{Results of exhaust gas analysis for three different gas mixtures. Mix~1 is 50.5\% Ar, 41.3\% C$_{2}$H$_{2}$F$_{4}$, 7.2\%i-C$_{4}$H$_{10}$ and 1\%SF$_{6}$, mix~2 is 80\% Ar and 20\% i-C$_{4}$H$_{10}$, with the addition of 1\% SF$_{6}$  and mix~3 is the same as mix~2, but without  SF$_{6}$. HV is the voltage applied to the RPCs, hit rate is the one measured on the RPC during the exhaust gas sampling and I is the current drawn.}
    \label{tab:table1}
    \begin{tabular}{c|c|c|c|c|c|c}
      \textbf{mix} & \textbf{HV} & \textbf{Hit rate} & \textbf{I} & \textbf{HF} & \textbf{HF/rate}& \textbf{HF/I}\\
       & kV & Hz/cm$^2$ & $\mu$A & $\mu$g/l $ $ & $ $\\
      \hline
      \#1 & 8.0 & 110 & 160 & 15 & $1.4\cdot10^{-4}$ & $0.9\cdot10^{-4}$ \\
      \#2 & 5.5 & 55 & 270 & 17 & $3.1\cdot10^{-4}$ & $0.6\cdot10^{-4}$ \\
      \#3 & 5.1 & 50 & 250 & 0 & $ 0 $ & $ 0 $ \\
      
    \end{tabular}
  \end{center}
\end{table}

Against our expectation, the amount of HF produced with the mixture with no C$_{2}$H$_{2}$F$_{4}$ at all was comparable with the one produced with the standard mixture with the same amount of SF$_{6}$ and a large amount of C$_{2}$H$_{2}$F$_{4}$: this shows that SF$_{6}$ plays a leading role in HF production. To understand the actual mechanism, further tests would have been needed, but the R\&D was halted to focus on the production and testing of the ALICE Muon Trigger apparatus. However, before the end of this study, we managed to perform one last, and crucial test, i.e. what is the ageing behaviour of an RPC operated with a gas mixture with no Fluorine at all?
To answer this question, we operated an RPC prototype with an 80\% Ar and 20\% i-C$_{4}$H$_{10}$ gas mixture (mix \#3 in Table\ref{tab:table1}) . Quite surprisingly, even if exposed to a rather low counting rate of $\sim$50~Hz/cm$^2$, the prototype showed the typical dark current increase, which characterizes ageing, reaching a dark current of tens of $\mu$A after about two weeks of operation.
The exhaust gas analysis of this RPC confirmed that no HF was present in the exhaust gas. This means that the HF is not the only factor causing dark current increase in the RPCs. While it is clear that the dark current grows with the integrated current, very different patterns of dark current increase emerged. Further tests will be needed to address the cause(s) of the dark current increase in RPCs.

\section{From streamer to maxi-avalanche}
\label{sec:maxiavalanche}

\begin{figure}[h!]
\center \includegraphics[height=0.6\linewidth]{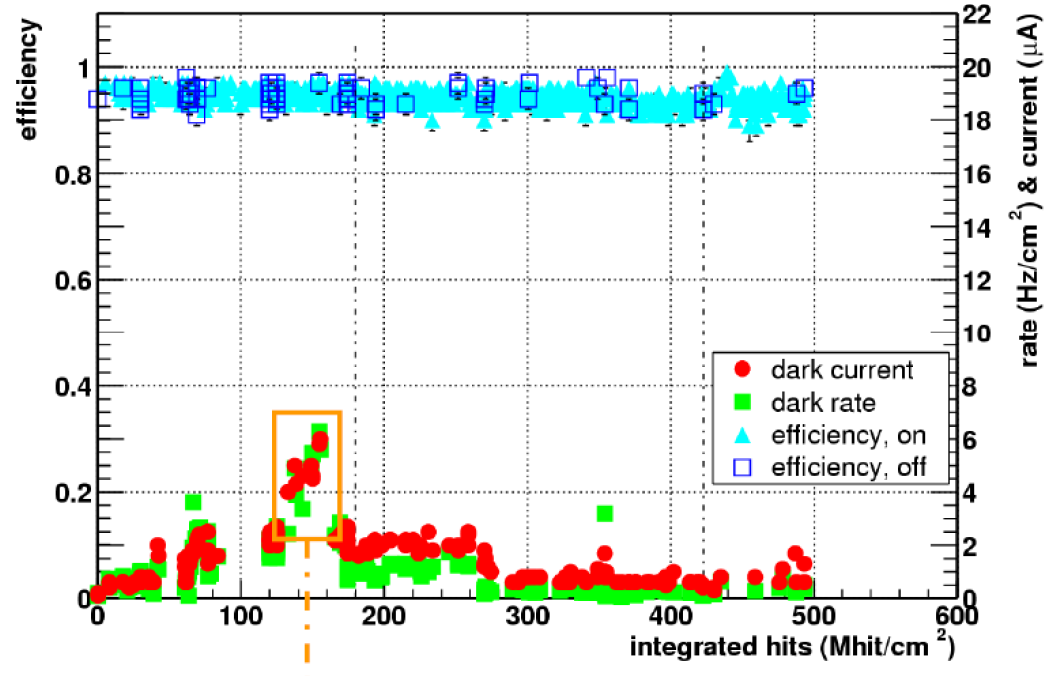}
\caption{Efficiency (left vertical axis), dark current and dark hit rate (right vertical axis) as a function of integrated charge measured in the final ageing test in 2004-2005. The temporary increase in current and rate at 150 Mhits/cm$^2$ highlighted in the orange box was due to a failure of the refrigerator, where the water bubbler used to humidify the gas mixture was kept.}
\label{fig:finalgif_2}
\end{figure}

The fundamental result of the tests was the correlation of ageing to the integrated charge. In ALICE, 100 Mhits/cm$^2 \cdot$ year were expected on the RPC detectors placed in the most exposed areas. To reach this goal, a strong reduction of the charge-per-hit was necessary. To minimize the signal charge, we adopted an ``avalanche-like''  (i.e. strongly quenched) gas mixture made of 89.7\% C$_{2}$H$_{2}$F$_{4}$, 10\% i-C$_{4}$H$_{10}$ and 0.3\% SF$_{6}$ (50\% relative humidity). However, since our readout electronics did not include a pre-amplification stage, we could not switch to the standard avalanche mode (the signal discrimination threshold is smaller than 1~mV). We decided to set the signal discrimination threshold to 10~mV, resulting in the so-called ``maxi-avalanche'' operation mode~\cite{maxiav}.
To verify if a RPC could withstand years of operation in ALICE, in 2004-2005 we performed a long-ageing test at GIF on a real-size RPC (270$\times$70 cm$^2$)\cite{maxiav2}. After accumulating 500 Mhits/cm$^2$  (corresponding to 100 mC/cm$^2$), the dark current was less than 1~$\mu$A and the dark-hit rate was less than 0.1~Hz/cm$^2$ (see Figure~\ref{fig:finalgif_2}).

\section{Ageing performance of the ALICE Muon TRigger/IDentifier}
\label{sec:maxiavalanche}

\begin{figure}[h!]
\center \includegraphics[height=0.6\linewidth]{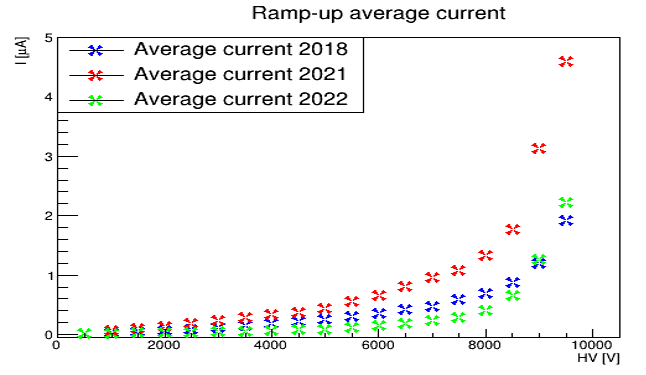}
\caption{Average currents measured during the HV ramp up of the Muon IDentifier. An increase can be noted in 2021 compared to 2018, even if the chambers were not operated during this period. After changing 10 HV cables, the average dark current at HV working point decreased from $\sim5\mu$A to $\sim2\mu$A}
\label{fig:avgcurr}
\end{figure}

According to the results of the ageing tests, and credit to a measured hit multiplicity in LHC Pb-Pb collisions which is significantly smaller than the one anticipated by simulations, it was decided to operate the Muon Trigger in maxi-avalanche mode, with a mixture composed of 89.7\% C$_{2}$H$_{2}$F$_{4}$, 10\% i-C$_{4}$H$_{10}$ and 0.3\% SF$_{6}$ and a detection threshold of 7~mV, in order to reduce as much as possible the charge per hit. In this configuration, the RPCs have been operating successfully since 2010. The most exposed detectors cumulated an integrated charge of $\sim$20~mC/cm$^2$ . The efficiency has  been constantly above 95\% all over the detector area. The main ageing issue has been due to the Polycarbonate gas outlets: some (about 10\%) broke and had to be replaced. In one case, a chamber had to be replaced because of an extended leak, probably due to bad glueing. Another issue was discovered in 2021 after the second LHC Long Shutdown (2018-2021). It was found that in some detectors, the dark currents were higher than at the beginning of the shutdown, despite the RPCs not being operated at all. The cause was in some HV cables and/or connectors leaking a small amount of current due to bad insulation. These cables were replaced during the 2021-2022 winter shutdown. In Figure~\ref{fig:avgcurr} the average dark currents measured during ramp-up before and after the replacement of cables are shown.

 Up to 2023, four RPCs were replaced because of relatively high dark currents at the HV working point ($\sim$30~$\mu$A), even if their efficiency was still uniform and above 95\%. In order to minimize the ageing, during the Long Shutdown, the ADULT front-end discriminators were replaced with FEERIC~\cite{feeric}, which includes a pre-amplification stage, allowing the RPCs to operate in pure avalanche mode, thus reducing the HV working point and the charge per hit.

%\vspace{-25pt}

\section{Conclusions and outlook}
\label{sec:conclusion}

RPCs of the ALICE MTR/MID behaved well thanks to an extensive ageing R\&D, which allowed us to understand potential issues well before the final installation and operation. 13 years after the start of operations, dark current and dark rate are under contro,l and the efficiency is high and uniform. Presently the RPCs are facing an environmental challenge: the C$_{2}$H$_{2}$F$_{4}$ and SF$_6$ gases are characterized by a high Global Warming Potential. Therefore, it would be advisable to replace them with low-GWP gases. TFor long-term operations, this challenge requires specific ageing tests to ascertain if the new gases do not cause bad effects. Work on this has already started, see Luca Quaglia's talk (included in these proceedings).

%\vspace{-18pt}

\end{document}